\documentclass[aps,pre,amsfonts,showpacs,unsortedaddress]{revtex4}
\usepackage[english]{babel}
\usepackage[latin1]{inputenc}
\usepackage[dvips]{graphicx}
\usepackage{amsmath}
\usepackage{amssymb}
\usepackage{epsfig}
\usepackage{array}

\newcommand{\be}{\begin{equation}}
\newcommand{\ee}{\end{equation}}
\newcommand{\Ref}[1]{(\ref{#1})}
\newcommand{\av}[1]{\left \langle #1\right \rangle}
\newcommand{\br}{{\bf r}}
\newcommand{\bv}{{\bf v}}
\newcommand{\bk}{{\bf k}}
\newcommand{\bn}{{\bf n}}

\begin{document}

\title{Power law tails of time correlations in a mesoscopic fluid model}
\author{M. Ripoll$^{a,b}$\thanks{Author to whom correspondence
should be addressed, e-mail:mripoll@fz-juelich.de} and M. H.
Ernst$^{c,d}$}
\affiliation{$(a)$ Institut f\"ur Festk\"orperforschung,
Forschungszentrum J\"ulich -  52425 J\"ulich, Germany\\
$(b)$ Dpto F\'{\i}sica Fundamental,
UNED, C/Senda del Rey 9, 28040 Madrid, Spain \\
$(c)$ CNLS, Los Alamos National
Laboratory, Los Alamos, NM 87545, USA \\
$(d)$ Institute for Theoretical Physics, Utrecht University,
Princetonplein 5, P.O. Box 80.195,  3508 TD  Utrecht, The
Netherlands}

\date{\today}

\begin{abstract}

In a quenched mesoscopic fluid, modelling transport processes at
high densities, we perform computer simulations of the single
particle energy autocorrelation function $C_e(t)$,  which is
essentially a return probability. This is done  to test the
predictions for power law tails, obtained from mode coupling
theory. We study both off and on-lattice systems  in one- and
two-dimensions. The predicted long time tail $\sim t^{-d/2}$ is in
excellent agreement with the results of computer simulations. We
also account for finite size effects, such that smaller systems are
fully covered by the present theory as well.
\end{abstract}


\maketitle

\label{sec:1}
\section{Introduction}
 In  classical fluids in thermal equilibrium the
velocity autocorrelation function (VACF) decays algebraically $\sim
t^{-d/2}$ at large times, and so do all Green-Kubo integrands, as
established by molecular dynamics
\cite{Alder+W,wood,Ashurst+levesque}, mode coupling
\cite{EHvL72,ME-univ}, and kinetic theory \cite{D-C-LTT}.
These functions are equilibrium time correlation function
$\av{J(t)J(0)}_0$ of single- or $N$-particle currents,
whose time integrals give the transport coefficients,
such as self-diffusion, viscosity or heat conductivity.

The long time tails are not restricted to current-current
correlations, but apply to large classes of time correlations,
such as single site correlations, provided the dynamics obeys a
conservation law. Similar single particle correlations with a long
time tail (LTT) exist in fluids \cite{ME-univ}. These tails have a
rather universal shape, in the sense that they are determined by
the decay of the slow macroscopic modes of the system, and are
independent of the microscopic details.

 Recently, we have discussed in paper I \cite{percol-I} a
simplified {\em mesoscopic} model for transport in fluids, in which
the rapid short range fluctuations are averaged out. It is referred
to as {\em random DPD solid} (Dissipative Particle Dynamics), and
defined by a fluctuating heat equation (Langevin equation), 
where the random force satisfies the fluctuation-dissipation
theorem. It is a special case of general DPD-fluids
\cite{percol-I,EW95}, in which {\it point} particles are
characterized by positions $\br_i$, velocities $\bv_i$ and possibly
by an internal energy $\epsilon_i$ $(i=1,2,\cdots,N)$.

The simplification has been obtained by quenching the
translational degrees of freedom of the DPD particles. The only
{\it dynamic} degrees of freedom left are the internal energies
$\epsilon_i  (i=1,2,\cdots,N)$ of the particles, where the total
energy $\sum_i \epsilon_i(t)=E$ is {\em conserved}. The only
remaining transport process is heat diffusion between fixed
particles, where energy hops instantaneously between {\em
interacting} pairs $\{ij\}$ within the interaction range
$(r_{ij}\le r_c)$. This transport mechanism is called {\em
collisional transfer}.
 The direction of the energy flow is determined
by the ``local temperature gradient''$(\epsilon_j-\epsilon_i)$,
and heat conduction is only possible for densities {\em above} a
{\em percolation} threshold $\rho_p$. In the underlying
percolating structure two particles are connected (by a bond)
if the distance between them satisfies, $r_{ij}\leq r_c$. So we
 are dealing with bond percolation
diffusion on the  random {\it proximity network}, where the
transition rate or conductivity  across a bond is {\it constant}.

In this paper we study the energy auto-correlation function of a
single particle $C_e(t)$.  It is the analog of the probability of
return to the origin, $P(R=0,t)$, where $P(R,t)$ satisfies a
diffusion equation.

The random DPD model is complementary to the Lorentz gas with
overlapping scatterers in several respects. The latter has only
{\em kinetic transport} of particles and the former has only the
transport mechanism of {\em collisional transfer} of energy. In classical
fluids both mechanisms are present. The former is {\em dominant}
at { low} and { moderate} densities and the latter at liquid
densities.

The overlapping Lorentz gas also has a {\em percolation
threshold}, where diffusion only occurs {\em below} the
percolation density. A more detailed comparison is given in paper
I, where the similarities and differences between both models for
percolation diffusion are discussed. The diffusion phenomena in
both models occur on random percolation structures, which are
essentially each other's duals \cite{percol-I}. In Lorentz gases
the return probability has a LTT $\sim t^{-d/2}$, and the VACF has
a LTT $\sim t^{-1-d/2}$.  In the hydrodynamic interpretation of
the LTT of the VACF in {\it classical fluids} in terms of
vorticity diffusion, as given in by Alder  and Wainwright, the
velocity auto-correlation function of a tagged particle may also
be interpreted as the return probability  of the initial momentum
to the tagged particle itself, where $d$ is the spatial dimension
of the system \cite{Alder-alley,Machta1,Machta2}. A further
property of the Lorentz gas is that the power law tail is {\em
absent} when the scatterers are placed on a {\em periodic} lattice
\cite{Harrison-Zwanzig}. The present paper also analyzes what
happens to the long time tails when the particles are put on a
lattice.  In fact,  the interpretation of $C_e(t)$, being a return
probability in a diffusion process, already suggests that $C_e(t)
\sim P(R=0,t)$ should also have a power law tail$\sim t^{-d/2}$ on a regular
lattice.

As the long wave length decay modes of classical fluids and DPD
models are essentially the same, the decay of the corresponding
{\em equilibrium} correlations should also be the same. This is in
fact implied by Onsager's regression hypothesis
\cite{book-deGroot-Mazur}. For the VACF in DPD fluids
\cite{esp99serr} and for the energy
autocorrelation function (EACF) in a DPD solid
\cite{Ripoll-thesis} more specialized mode coupling arguments have
been developed to show the existence of a power law tail $\sim
t^{-d/2}$. Unfortunately the predictions, in particular the one
for the DPD solid, could not be confirmed by the existing computer
simulations of three-dimensional DPD systems, because the systems
studied were too small, e.g. $N=1000, L/r_c =5$ and $N=2000, L/r_c
=6$, where $V=L^d$ \cite{Ripoll-thesis}. Consequently, the short
time kinetic decay, $\exp[-\omega_0 t]$, and the slow decay of the
macroscopic diffusive modes, $\exp[-k^2 Dt]$, were equally fast
(no separation of time scales), and power law tails in the EACF
could {\it not} be observed because of strong finite size effects.
The system sizes in our present computer simulations of the
two-dimensional DPD solid are sufficiently large to observe the
power law tails.

The plan of the paper is as follows. In section II the definitions
and results for the DPD solid are briefly summarized in so far as
needed in the present paper. In section III the mode coupling
theory of \cite{EHvL72,ME-univ} is applied to obtain the explicit
form of the LTT of $C_e(t)$ in the random DPD solid.   In section
IV the results are compared with computer simulations of the
two-dimensional case. In section IV it is also explained how the
{\it same} results for the time correlations can be obtained from
deterministic simulations, where the dynamics is {\it free of
statistical noise}. In section V the LTT's of time correlation
functions are studied on a lattice. Conclusions are presented in
section VI.

\section{Random DPD Solid}

\label{sec:2}
The system consists of $N = n V$ {\em point} particles at fixed
random positions $\br_i (i=1,2,\ldots,N)$, contained in a volume
$V=L^d$, and obeying periodic boundary conditions. Each DPD
particle interacts with all particles inside its interaction
sphere of radius $r_c$. The only dynamical variables in the system
are the internal energies $\epsilon_i$ of the particles, and the
time evolution is given by the Langevin equation,
\be \label{2.1}
\frac{d\epsilon_i}{dt}=\lambda_0 {\sum_j}^\prime w(r_{ij})
\left( \epsilon_j-\epsilon_i \right)
+ {\sum_j}^\prime \widetilde{F}_{ij}(t).
\ee
Here $w(r)$ is a positive interaction function of finite range
$r_c$, normalized as $\int d\br w(r)=r_c^d$. In the present paper
$w(r)= const \: \theta(r_c-r)$ is  proportional to the unit step
function,  which vanishes for $r>r_c$. The kinetic rate constant
$\lambda_0$ is a model parameter that determines the interaction
frequency. The prime on the summation sign indicates that $j\neq
i$. Here $\widetilde{F}_{ij} = -\widetilde{F}_{ji}$ is Gaussian
white noise, whose explicit form is given in paper I. It satisfies
the detailed balance conditions, guaranteeing that the system
reaches {\em thermal equilibrium}.

To discuss the short time decay of an energy fluctuation we
consider
\be \label{2.3}
{\nu} = {\sum_j}^\prime\av{\theta(r_c-r_{ij})}\equiv \rho \varpi_d,
\ee
where $\nu$ is the mean number of $j$-particles inside $r_{ij}\le
r_c$ that interact with the $i$-th particle. Moreover  $\varpi_d =
\pi^{d/2} /\Gamma (1+d/2)$ is the volume of  a $d-$dimensional
unit sphere $(d=1,2,\cdots)$, and $\rho = n r_c^d$ is the reduced
density.

The average $\av{\cdots}$ denotes a quenched
average over the random configurations of the fixed DPD particles.
The basic relaxation rate at short times can be estimated from
\Ref{2.1} and
\Ref{2.3} as
\be \label{2.3a}
\omega_0 = \lambda_0 {\sum_j}^\prime \av{w(r_{ij})} =   \rho \lambda_0.
\ee
Each DPD particle is a mesoscopic subsystem with a density of
internal states $\sim \epsilon^\alpha$, where $\alpha$ is
proportional to the number of internal degrees of freedom of a DPD
particle, and satisfies $\alpha \gg 1$. In thermal equilibrium the
single particle distribution function is
\begin{equation} \label{2.4}
\psi_0(\epsilon) =
\frac{\beta}{\Gamma(\alpha+1)}\left(\beta\epsilon\right)^\alpha
\exp[-\beta\epsilon],
\end{equation}
where $T=1/k_B \beta$ is the temperature. Also note that the
evolution equation conserves the total energy, i.e. $\sum_i
\epsilon_i (t) = E = const$, as both terms in \Ref{2.1} are
anti-symmetric in $i$ and $j$.

A final comment regarding the stochastic differential equation
\Ref{2.1}, which contains multiplicative noise. As discussed
in paper I, the difference between the $\hat{\rm{I} }$to and
Stratonovich interpretation of the Langevin equation and the
corresponding Fokker Planck equation can be
neglected to leading order in ${\cal O}(1/\alpha)$ in the present
model.

In this paper we study the long time tail of the energy
autocorrelation function (EACF) in thermal equilibrium,
\be \label{2.7a}
C_e(t) = \frac{1}{N} \sum_i \av{\delta \epsilon_i(t)\delta
\epsilon_i(0)}_0,
\ee
where $\av{ \cdots}_0$ is an average over the canonical
distribution $\prod_j \psi(\epsilon_j)$. Here
$\delta\epsilon_j(t)=\epsilon_j(t)-\av{\epsilon_j}_0$ is the
energy fluctuation with $\av{\epsilon_j}_0 \simeq \alpha/\beta$. It
is essentially the return probability of the initial energy $
\delta
\epsilon_i(0)$ to the particle, from which it originated. It is the
analog of the velocity autocorrelation function,  $C_v(t) =
\av{v_x(t)v_x(0)}_0$ in the same sense as the VACF is the return
probability of the initial momentum. The latter picture explains
its LTT $\sim t^{-d/2}$
\cite{Alder+W}.
However the time integral of the VACF equals the coefficient
of self diffusion $D$, whereas the time integral of the EACF
is not related to the heat diffusivity $D_T$, or to any
other transport coefficient.

The explicit form of the power law tail of the EACF has been given
in Refs.\cite{Ripoll-thesis,ME-univ}. The derivation can
essentially be copied from the corresponding derivation for the
VACF in fluids \cite{EHvL72}. We only give an outline. We first
express $\delta \epsilon_1 (t) = \int d \br \delta e_s(r,t)$ where
$\delta {e}_s(r,t) = \delta \epsilon_1(t)
\delta(\br -\br_1)$ is the local energy density of tagged
particles. According to the Onsager regression hypothesis, the
average decay of fluctuations around thermal equilibrium follows
the macroscopic approach to equilibrium. This implies that a local
fluctuation $\delta e_s$ {\it rapidly} decays to its value in
local equilibrium, i.e.
\be \label{3.2}
\delta \epsilon_1 (t) =\int d\br \delta {e}_s (r,t) \simeq \int d\br
n_s(r) {\cal C} \delta T (r,t) \simeq \frac{{\cal C}}{V} \sum_{\bf
k}
\delta T_{\bf k} (t) n_{-\bf k}^s,
\ee
where ${\cal C} =\alpha k_B$ is the specific heat per DPD
particle, and $\delta T(\br,t)$ is the local temperature
fluctuation, $n_s(\br) = \delta (\br -\br_1)$ the quenched local
density of tagged particles, and $a_{\bf k}$ denotes the Fourier
transform of $a(\br)$. The subsequent {\it slow} evolution is
controlled by the heat mode $\delta T_{\bf k}(t) = \delta T_{\bf
k}(0) \exp [-k^2D_T t]$.  The tagged particle density $n^s_{\bf
k}$ is a static mode. Inserting $\delta T_k$ in \Ref{2.7a} and
\Ref{3.2}, and using the relations $n_k^s \to 1$ as $k \to 0$, and
$\delta T_k(0) \to \delta \epsilon_1(0)/{\cal C}$ as $k \to 0$, we
obtain in the {\it long time limit},
\be \label{3.3}
C_e(t) =  \frac{1}{nV} \sum_{\bf k} \exp[-k^2 D_T t]C_e(0).
\ee
This yields in the {thermodynamic limit} as $N=nV \to \infty$ at
fixed $n$, the { power law tail},
\be \label{3.4}
\frac{C_e(t)}{C_e(0)} \to \frac{1}{n} \int \frac{d {\bf
k}}{(2\pi)^d} \exp[-k^2 D_T t] = \frac{1}{n(4\pi D_T t)^{d/2}} =
\frac{1}{\rho\left(4\pi t^*\right)^{d/2}} \qquad \left( t^* \gg 1
\right)
\ee
In the first step we used the relation $C_e(0) = \av{( \delta
\epsilon )^2}_0 \simeq \alpha /\beta^2$, and
in the last we introduced the dimensionless time $t^*=D_T t /
r_c^2$, where $t_D = r_c^2 / D_T(\rho)$ is the characteristic time
for heat diffusion. The integral representation \Ref{3.4}, $\int
d\bk P_\bk(t) \sim P(R=0,t)= (4 \pi D_T t)^{-d/2}$, of the long
time limit of $C_e(t)$ shows that $P(R,t)$ is the solution of the
heat diffusion equation, and that $P(R=0,t)$ is a return
probability.

 The main goal of this paper is to test the theoretical prediction
 about the long time tail (LTT) in $C_e(t)$.

\section{Long time tails in stochastic and deterministic simulations}
\label{sec:4}

The goal of this section is to compare the predictions
\Ref{3.3} for the long time tails with the results of
computer simulations in the two-dimensional random solid described
by the Langevin equation \Ref{2.1}. The values of the heat
diffusivity, to be used in the comparison, have been obtained from
computer simulations, using the methods of paper I, and agree well
with the mean field result $D_\infty (\rho) =
\lambda_0 \rho r_c^2 / [2(d+2)]$ at {\it high densities}.
 As $\rho$ decreases, $D_T(\rho)$ {\it decreases faster}
than linear, and vanishes at a non vanishing threshold density
$\rho_p$, which coincides with the percolation threshold for
continuum percolation of overlapping spheres. In two dimensions it
has the value $\rho_p
\simeq 1.43629$ \cite{ziff}, and in three dimensions one has $\rho_p=0.65296$
\cite{ziff-3d}.

To measure LTT's of {\it equilibrium} time correlation functions
the DPD solid has to be in a state of thermal equilibrium. This is
achieved by initializing the system in a state with a uniform
energy distribution $\epsilon_i(0)=E/N$, where $E$ is the total
energy. In principle the Langevin forces, which satisfy the
detailed balance conditions, drive the system to a state of
thermal equilibrium, described by $\psi_0$ in Eq.\Ref{2.4}. This
is only the case in a pure ergodic system. However with
decreasing  density larger fractions of  particles are contained
in small disconnected islands, which have no interactions with the
bulk of the particles in the system, i.e. belong to small
independent ergodic subsystems, and cannot redistribute their
initial energies over the bulk system. Hence, or initialization of
the system in the canonical equilibrium state can only be
guaranteed if the fraction of island particles is negligible, i.e.
for $\rho \gg
\rho_p $.

%{\bf Figure 1.}
\begin{figure}[h]
\includegraphics[width=5.5cm,angle=-90]{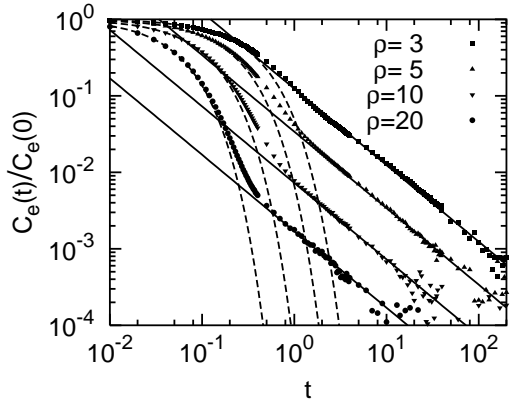}
\caption{Simulation data for EACF, $C_e(t)$
versus $t$ at different reduced densities $\rho$ (defined below
Eq.\protect{\Ref{2.3}}) in the two-dimensional random solid for a
system of $N=\rho(L/r_c)^2=5\times 10^4$ particles, compared with
theoretical predictions: the dashed lines represent the short time
exponential decay, and the solid lines the algebraic LTT$\sim
1/t$. The higher the density the sooner the LTT is reached.}
\end{figure}

%{\bf Figure 2.}
\begin{figure}[h]
\includegraphics[width=5.5cm,angle=-90]{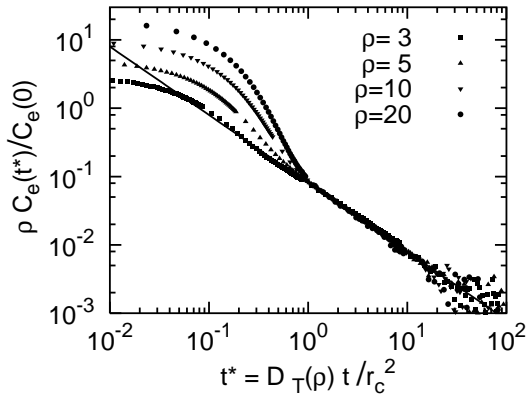}
\caption{Collapse plot of the LTT's in
Fig.1, obtained by plotting $\rho C_e(t)/C_e(0)$ in
Eq.\protect{\Ref{2.7a}} versus $t^*$, compared with the predicted
LTT (solid line). Note that the crossover from exponential to
algebraic decay is at $t_D=r^2_c/D_T(\rho)$.}
\end{figure}

%{\bf Figure 3.}
\begin{figure}[h]
\includegraphics[width=5.5cm,angle=-90]{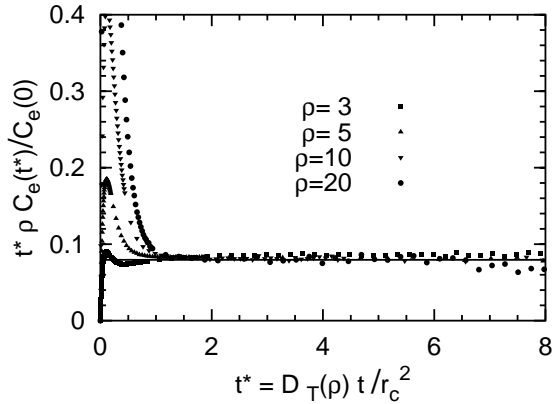}
\caption{A more sensitive comparison of the
collapse plot with the theoretical prediction (solid
horizontal line), focusing on relatively short times.}
\end{figure}

Figs. 1,2,3 show the simulation data for the energy
auto-correlation function $C_e(t)$ of a DPD particle in a
two-dimensional system of $N=\rho (L/r_c)^2$ particles at
different densities. There is excellent agreement of computer
simulations with theoretical predictions. In the simulations we
measure length in units of $r_c$, and times in units of
$1/\lambda_0$.  At {\it short} times the decay is {\it
exponential}, $\exp[-\omega_0 t]$, with a rate constant $\omega_0
= \rho \lambda_0$, as given in
\Ref{2.3a}, in very good agreement with the simulation data. At
{\it large} times the plots show the long time tail $\sim
t^{-d/2}$.

By plotting the simulation results as $\rho C_e(t) / C_e(0)$
versus the dimensionless time $t^* = D_T(\rho)t/r_c^2$ the data
for different $\rho$ can be collapsed on a single LTT-curve.
Moreover the combination of Figs.1, 2 and 3 shows that the
crossover time from {\it exponential} short time decay $\sim
\exp[-\omega_0 t]$ to {\it power law} decay is for all densities
given by $t_D=r_c^2/D_T(\rho)$.

The finite size effects on the dynamics are controlled by the
ratio $L/r_c=(N/\rho)^{1/d}$.  As DPD particles in absence of
conservative forces are {\it point} particles, the density can
become arbitrarily high. So, as $\rho$ increases at fixed $N$ the
finite size effects increase.  Consider the data in Figs.1, 2 and
3 at $(\rho=5, L=100)$, corresponding to $N=5\times10^4$
particles. If the number of particles is reduced to $N=10^4$ a
faster decay than $1/t$ decay becomes noticeable, which is
statistically significant. For $N=10^3$ the decay is even faster,
and looks {\it exponential}. It is caused by finite size effects.
We return to this point later on.

The stochastic simulations above were described by the Langevin
equation \Ref{2.1}. The subsequent deterministic simulations of
this section correspond to the same equation with the Langevin
forces switched off ($\tilde{F}_{ij}(t)=0$).   Moreover, the
stochastic forces have no effect on the decay of the {\it
equilibrium} time correlation functions $\av{\delta a(t)
\delta a(0)}_0$, provided both types of correlation functions
are at $t=0$ in thermal
equilibrium. This observation follows from {\it Onsager's
regression hypothesis} on the average decay of fluctuations. The
purpose of driving the system by Gaussian white noise, that
satisfies the detailed balance conditions, is to maintain the
system in thermal equilibrium, but does not affect the decay of
these functions.

The above observations offer the interesting possibility to
measure the LTT of equilibrium time correlations
deterministically, i.e. the dynamics is free of {\it statistical
noise}, provided the system is prepared initially in the proper
thermal equilibrium state. Here the thermal fluctuations are only
accounted for in the initial distribution.
 This method has also been used in \cite{LBE} 
 where  the kinetic part of the stress-stress correlation was
 measured using the lattice Boltzmann equation.

Next we discuss the results of the {\it deterministic} simulations
of $C_e(t)$ in the two-dimensional random DPD solid at density
$\rho=5$, where on average $\pi\rho$ particles are surrounding the
central particle in the interaction sphere. In order to simulate
the equilibrium time correlation function $C_e(t)$, we prepare the
system in the initial state, described by the $N-$particle
distribution, $\prod^N_{i=1} [\psi_0(\epsilon_i)/V]$, and average
over different runs with independent initial configurations. In
spite of the additional average over different runs the
computational effort for the deterministic simulations is
considerably smaller than  for the stochastic ones. Figs. 4 and 5
show log-log plots of $C_e(t)/C_e(0)$ versus $t$ at different
system sizes $L/r_c$ with $N=\rho(L/r_c)^d$ particles. Fig. 4 is
focusing on  the {\it short} time behavior $\exp[-\lambda_0 \rho
t]$ and {\it power law} tails $(1/\rho)[4 \pi D_T(\rho)
t/r_c^2]^{-d/2}$, and Fig. 5 on the {\it power law} tail and the
ultimate {\it exponential decay}. Both figures also show for
comparison the stochastic simulation at $L/r_c = 44$  as scattered
black circles. This confirms that the exponential short time and
the intermediate power law behavior of the deterministic and
stochastic simulations are the same, and agree with the
theoretical prediction, until  the time where the statistical
uncertainty in the stochastic simulations has become too large.
This behavior beautifully confirms Onsager's regression hypothesis.

%{\bf Figure 4.}
\begin{figure}[h]
\includegraphics[width=5.5cm,angle=-90]{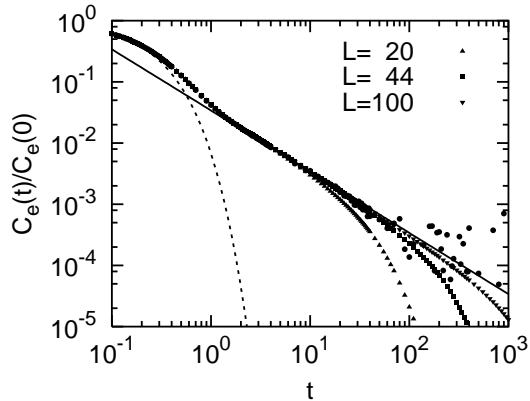}
\caption{The log-log plot shows the EACF vs $t$ in the deterministic
simulations at $\rho=5$ for $L/r_c=20,40,100$, corresponding to
$N=2\times 10^3, 8\times 10^3, 5\times 10^4$ particles. The scattered
black circles ($L/r_c=44$) show stochastic simulations for
comparison. The straight solid line shows the predicted power law tail
\protect{\Ref{3.4}}, and the dashed line indicates the short time
exponential decay.}
\end{figure}

%{\bf Figure 5.}
\begin{figure}[h]
\includegraphics[width=5.5cm,angle=-90]{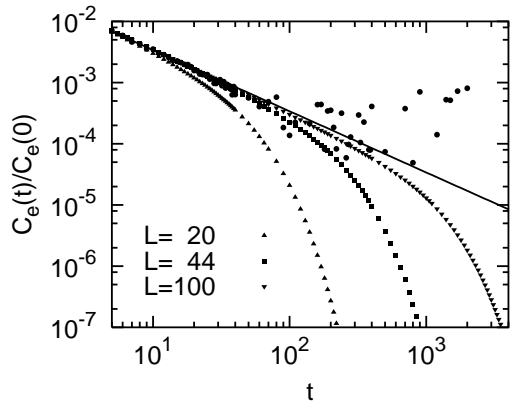}
\caption{The log-log plot shows the same
deterministic simulations as in Fig.4, but focuses on the
intermediate algebraic tail. The straight line is the LTT in
the thermodynamic limit. The ultimate long time decay is again
exponential. For the scattered black circles see caption Fig.4.}
\end{figure}

%{\bf Figure 6.}
\begin{figure}[h]
\includegraphics[width=5.5cm,angle=-90]{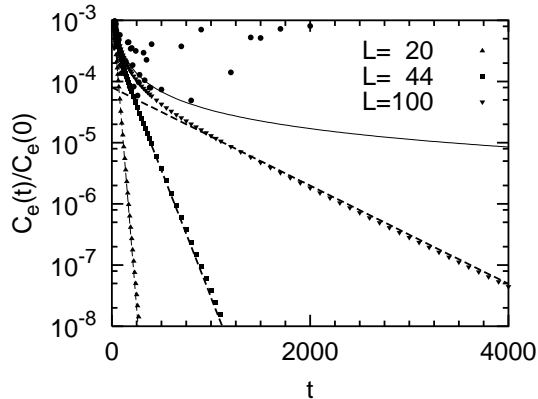}
\caption{The half-logarithmic plot focuses
on the ultimate exponential decay in the simulations of Figs. 4
and 5, which are finite size effects, and are quantitatively
accounted for by the first term in Eq.\protect{\Ref{4.1}},
$(4/N)\exp[-(2\pi/L)^2D_Tt]$. The solid line at the top right is
again the algebraic LTT for thermodynamically large systems.
For the scattered black circles, see caption Fig.4.}
\end{figure}

Fig. 6 shows that the ultimate decay is again exponential. This is
a finite size effect, which is in excellent agreement with the {
mode coupling prediction} \Ref{3.3} for {\it finite} systems,
where the $\bk$-summation can {\it not} be replaced  by a
$\bk$-integral. On a square lattice with periodic boundary
conditions $k_\alpha = 2 \pi n_\alpha/L$ with $\alpha = x,y$ and
$n_\alpha = 0, \pm1, \pm2, \cdots$. The integral approximation to
\Ref{3.3} is only valid in the {\it limit} of {\it large} $t$, and
{\it small} $k$, such that $k^2 t =$ constant. This asymptotic
approximation is appropriate as long as $D_T k_{min}^2 t \ll 1$,
and breaks down at $t_{min}$ for, say, $D_T k^2_{min}t_{min}
\approx 1/3$, where the minimum $k$-value is $k_{min} = 2 \pi /L$.
The first few terms in the finite size mode coupling formula on a
square lattice follow from \Ref{3.3} in the form,
\be \label{4.1}
\frac{C_e(t)}{C_e(0)} = \frac{1}{N} \sum_{\bf n}
\exp[-|\bn|^2 s] = \frac{1}{N} \left( 4 e^{-s} + 4 e^{- 2
s} +4 e^{- 4 s} +8 e^{- 5 s} + \cdots \right),
\ee
where  $s = k^2_{min} D_T t = 4 \pi^2 D_T t / L^2$, and ${\bf n}
=(n_x,n_y)$. In Fig. 6 the straight lines of simulation data are
well represented through the first term $(4/N)e^{-s}$ of the right
hand side of \Ref{4.1}.

Furthermore, our  method is not suitable to study equilibrium time
correlations close to the percolation, because the  equilibration
time $t_0$ diverges as $\rho \downarrow \rho_p$  on the
percolating cluster. Moreover the finite fraction of
particles contained in smaller disconnected islands form uncoupled
ergodic subsystems. This prevents the system from  reaching the
canonical equilibrium state, $\prod_j\psi_0(\epsilon_j)$.

\section{DPD - solid on a lattice}
\label{sec:6}

In this section we introduce a lattice version of the DPD solid in
\Ref{2.1}, where $N$ particles fill the $N$ sites of a hypercubic
lattice with lattice distance ${a}$, and volume $V=L^d=Na^d$. In
this model the interaction range $r_c$ is the control parameter
and heat diffusion only occurs for $r_c\geq a$. We will observe
that the LTT of $C_e(t)$ in the random DPD solid survives when the
DPD particles are put on a periodic lattice, as is expected for a
return probability. To do so we consider the special case of the
DPD solid on a lattice, where $C_e(t)$ can be calculated exactly
\cite{ME-univ}. However, in periodic Lorentz gases the power
law tail disappears in the VACF, as shown in
Ref.\cite{Harrison-Zwanzig}.

Consider the evolution equation for the lattice {\it space-time}
correlation function $C(\bn,t)=\langle \delta
\epsilon_\bn(t)\delta\epsilon_{\bf 0}(0)\rangle _0$ in thermal
equilibrium, where $\bn=(n_x,n_y,\dots,n_d)$ labels the sites of a
hypercubic lattice with periodic boundary conditions, and
$n_\alpha=\{0,1,2,\dots,L-1\}$ with $\alpha=\{x,y,\dots,d\}$. The
evolution equations are obtained by replacing $\epsilon_i$ in
\Ref{2.1} with $\delta \epsilon_\bn(t)$, multiplying that equation
with $\delta \epsilon_{\bf 0}(0)$, and averaging over the
$N$-particle equilibrium distribution function.  As $\delta
\epsilon_{\bf 0}(0)$ is uncorrelated with the Langevin force, the
equations of motion of the lattice space-time correlation
functions follow as,
\begin{equation}\label{6.1}
dC(\bn,t)/dt =\lambda_0 \sum_{\bf m} w(r_{\bf nm}) \left(C({\bf
m},t) -C(\bn,t)\right)
\end{equation}
with initial condition $C(\bn,0)=\delta_{\bf n0} \langle(\delta
\epsilon)^2\rangle_0 \simeq \delta_{\bf n0} \alpha/\beta^2$. This is
a discrete diffusion equation, which can be solved exactly for
general dimensionality \cite{ME-univ}, and the single site
correlation or return probability  $C({\bf n= 0},t) =C_e(t)$. For
$w(r_{\bf nm})$ restricted to nearest neighbor interactions $(r_c=a)$, 
the solution is given by
\cite{ME-univ},
\begin{equation}\label{6.2}
C_{e}(t)/C_{e}(0)=\left[e^{-\tau}I_0(\tau)\right]^d \to
(2\pi\tau)^{-d/2}.
\end{equation}
Here $\to$ denotes the long time behavior for $\tau\gg 1$.
Moreover $I_l(\tau)$ with $l=0$ is a modified Bessel function with
$\tau=2D_T t/a^2$, and the heat diffusivity, $D_T=\lambda_0
a^2/\varpi_d$, where units of time are consistent with \Ref{2.1}
and \Ref{2.3a}.

In the lattice model both characteristic time scales, the short
kinetic time scale, $1/\omega_0=1/\rho\lambda_0$, and the long
diffusive time scale $t_D=r_c^2/D_T(\rho)$, are roughly inversely
proportional to the mean number of $j$-particles inside the
interaction sphere $(r_{ij}\leq r_c)$ of the central particle $i$, 
and differ roughly by an order of magnitude in size 
As $\rho\downarrow\rho_p$ this number decreases rapidly, and the
equilibration time $t_0$ at $\rho_p$ diverges, as the interactions
become weaker and weaker.

For the exactly solved case of the lattice DPD solid in
$d$-dimensions with nearest neighbor interactions, Figs. 7 and 8
show for the one-dimensional case that a very long
equilibration time $\lambda_0t_{0}\simeq 5\times 10^5$ is required.
After that period the exact solution $C_{e}(t)$ can be observed
with its full $1/\sqrt{t}-$tail over the whole time interval
$\lambda_0t\leq 1000$. If the equilibration time is too short, the
observed LTT falls off faster than $1/\sqrt{t}$ .

%{\bf Figure 7.}
\begin{figure}[h]
\includegraphics[width=5.5cm,angle=-90]{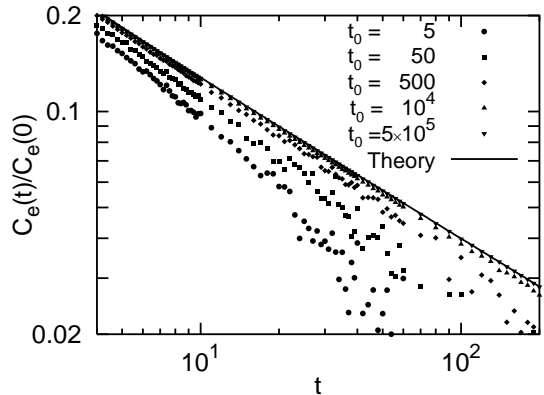}
\caption{The plot shows the simulated EACF
on a one-dimensional lattice with nearest neighbor
interactions after different equilibration times
$t_0=5,50,500,10^4$ and $5\times 10^5$, as well as the exact
solution \protect{\Ref{6.2}} for the one-dimensional
lattice model.}
\end{figure}

%{\bf Figure 8.}
\begin{figure}[h]
\includegraphics[width=5.5cm,angle=-90]{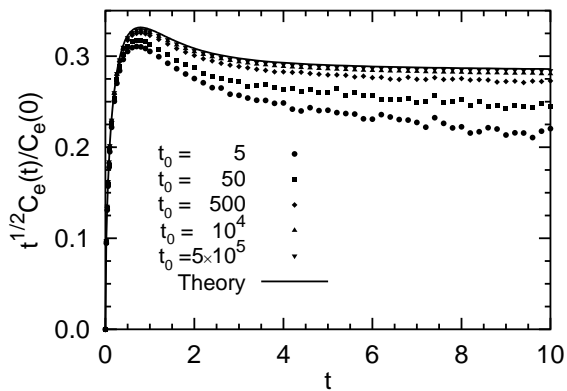}
\caption{The plot shows the same data for
$C_e(t)$ as in Fig.7, but focuses on short times. Note that the
interval around $\lambda_0 t\simeq 1$ also equilibrates
very slowly. }
\end{figure}

For general interaction ranges $(r_c= {\cal N} a)$, where ${\cal
N}$ represents the number of interacting lattice sites  on
periodic lattices, the mode couplings arguments of
\cite{Ripoll-thesis,ME-univ} produce the LTT in \Ref{3.4} which yields in
one dimension an algebraic LTT $\sim 1/\sqrt t$. In the
one-dimensional lattice case it is straightforward to calculate
the coefficient of heat diffusivity $D_T(\rho)$ analytically, as
shown in the appendix, and the results of the LTT's in the
stochastic simulations are show in Figs. 9, 10, 11, which can be
extended directly to higher dimensional lattices. The simulation
results are in very good agreement with the predictions of the
mode coupling theory.

%{\bf Figure 9.}
\begin{figure}[h]
\includegraphics[width=5.5cm,angle=-90]{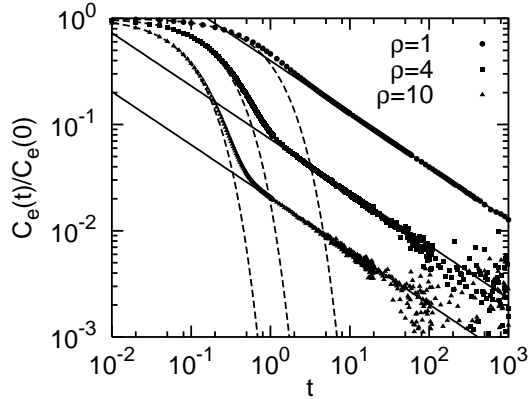}
\caption{The EACF $C_e(t)$ in the one-dimensional
lattice model at different $\rho$ for a system of $N=10^4$
particles compared with the theoretical predictions: dashed line is
the short time exponential decay, and the straight lines are the
algebraic LTT $\sim 1/\sqrt t$ (Compare with Fig.1).}
\end{figure}

%{\bf Figure 10.}
\begin{figure}[h]
\includegraphics[width=5.5cm,angle=-90]{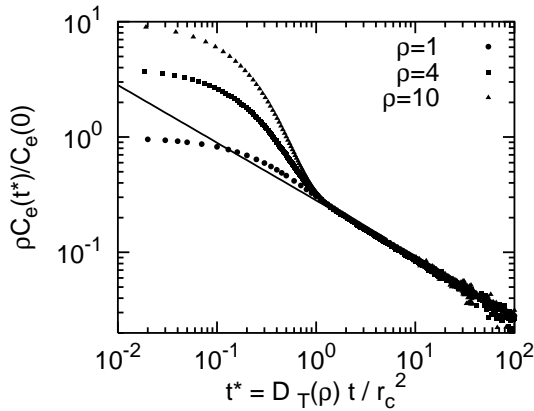}
\caption{Collapse plot of the LTT in the one-dimensional
lattice model (Compare with Fig.2).}
\end{figure}

%{\bf Figure 11.}
\begin{figure}[h]
\includegraphics[width=5.5cm,angle=-90]{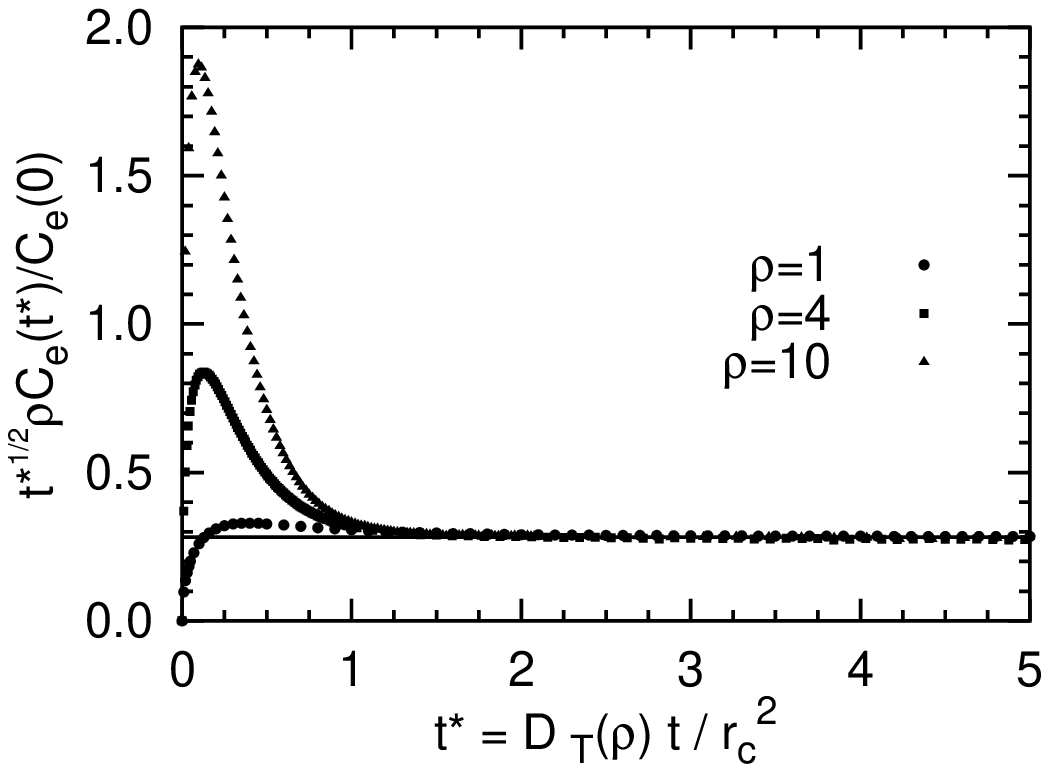}
\caption{A more sensitive comparison of the
collapse plot of Fig.10 for short times
(compare with Fig.3) with the theoretical predictions.}
\end{figure}

We also note that the {\it random} DPD model in one dimension 
is in the thermodynamic limit nonconducting because at any
given density $\rho$ there will always be a non-vanishing
probability to find a single pair $(ij)$ of nearest neighbors with
$r_{ij}> r_c$.

\section{Conclusions and Open Problems}
\label{sec:7}

The general conclusion from the previous sections is that the
predictions of mode coupling theory for classical fluids
\cite{EHvL72} regarding the existence of LTT $\sim t^{-d/2}$ in
time correlations, are in excellent agreement with the results of
computer  simulations, when applied to the energy autocorrelation
function $C_e(t)= \langle \delta \epsilon_i(t) \delta
\epsilon_i(0)\rangle_0$ in the random DPD solid and in the lattice
DPD models (with a uniform density distribution) both in one and
two dimensions. Here the energy auto-correlation function $C_e(t)$
at large time is essentially a return probability in the same
sense as the velocity auto-correlation function is in a classical
fluid.

Nevertheless, a number of interesting questions about consistency
of the theory remains for the LTT-predictions  $C_e(t) \propto (D_T
t)^{-d/2}$ for $d=1,2,\dots$. For classical fluids in two
dimensions the Navier Stokes transport coefficients do not exist.
For example, let $C(t)=\langle v_x(t)v_x(0)\rangle_0$ with a LTT
$\sim t^{-d/2}$, be the velocity autocorrelation function. Then the
long time limit of $D(t)\equiv \int_0^t d\tau C(\tau)$ yields the
coefficient of self diffusion $D$, if the integral exists. However,
this relation would lead to the result $D(t)\sim \{\sqrt t,\ln t\}$
for  $d=\{1,2\}$ respectively in the long time limit, leading to
the non-existence of the Navier- Stokes transport coefficients in
one-dimensional and two-dimensional fluids. Then, self
consistent mode coupling theory, ring kinetic theory and other
renormalization procedures \cite{ME-physicaD} lead to a
"renormalized super-long time tail" of the form
\cite{Fr-superLTT,ME-physicaD},
\begin{eqnarray}
D(t)&\sim & \{t^{1/3},\sqrt{\ln t}\}\nonumber \\
C(t) &\sim & dD(t)/dt =\{ t^{-2/3},1 / [t\sqrt{\ln t}]\}
\label{7.1}
\end{eqnarray}
for $d=\{1,2\}$ respectively. In fact, van der Hoef and Frenkel
\cite{Fr-superLTT} have confirmed for the VACF in two-dimensional fluids the
existence of a "faster-than-$t^{-1}$-tail" of $C(t)$ by computer
simulations using lattice gas cellular automata. However, our
computer simulations of $C_e(t)$ in Figs.1-3 and 9-11 do not show
any deviations from the predictions of the simple mode coupling
theory in \Ref{3.4}, and seem to confirm the LTT $\sim t^{-d/2}$,
as well as the finiteness of $D_T$.

This simple behavior is further confirmed by the observations (i)
that $C_e(t)$ in the two-dimensional  random solid in Figs.1, 2, 3
and in the one-dimensional lattice versions in Figs. 9, 10, 11
behave essentially the same, and (ii) that the lattice models
contain as a special case the model with  nearest neighbor
interactions (with $r_c=a$), which is {\it exactly soluble} for
all values of $d$
\cite{ME-univ}, and show the same LTT $\propto (D_Tt)^{-d/2}$ with
a {\it finite }  $D_T$, as the random DPD solid and the remaining
lattice models with $r_c=Ma$, where $M=1,2,\dots$. In fact, the
behavior of the DPD solid resembles that of Lorentz gases, where
the diffusivity $D$ is also finite for all values of $d$. In fact
the random DPD solid and the Lorentz gas are in several respects
each others duals \cite{percol-I}, where the VACF has the  LTT
$\sim t^{-1-d/2}$.

Furthermore, there is no contradiction between the long time
behavior of fluids in \Ref{7.1} and that of the DPD solid in
\Ref{3.4}, because $D_T$ in the latter case is {\it not} related
to the time integral of $C_e(t)$, but is given by the Green-Kubo
formula, involving the time correlation function $C_Q(t)$  of the
microscopic heat current,
 whose time integral converges \cite{ME-univ}.

\section*{Acknowledgments}
The authors thank E. Ben-Naim, J. Machta, R.W. Ziff,  G.
Vliegenthart, P. Krapivsky, R. Brito, and P. Espa\~nol for helpful
discussions and correspondence. M.R. also acknowledges financial
support from the Ministerio de Ciencia y Tecnolog{\'{\i}}a under
the project BFM2001-0290 and the German Research Foundation (DFG)
within the SFB TR6, ``Physics of Colloidal 
Dispersions in External Fields''.

\appendix

%%%%%%%%%%%%%%%%%%%%%%%%%%%%%%%%%%%%%%%%%%%%%%%%%%%%%%%%%%%%%%
\renewcommand{\theequation}{A.\arabic{equation}}
\setcounter{equation}{0}
\label{app:a}
\section{}

In this appendix we calculate the coefficient of the {\it heat
diffusion} for the lattice version of the DPD model in one
dimension with an interaction range $r_c = M a$ with
$M=1,2,\cdots$ for the evolution equation \Ref{2.1} with the
Langevin force set equal zero, and using  the range function $w(r)
= \frac{1}{2} \theta(r_c - r)$ with $[w]=r_c$ for the
one-dimensional case. The evolution equation \Ref{2.1} for
$r_{nm}\geq Ma$ becomes,
\be \label{a1}
\frac{\partial \epsilon(r_n,t)}{\partial t} = \frac{1}{2} \lambda_o
\sum_{|m|\le M} \left[ \epsilon (r_m+ma,t) - \epsilon(r_n,t)\right],
\ee
where $M=\pm 1$ for nearest neighbors. For large spatial scales,
this discrete version of the diffusion equation is expanded to
${\cal O} (\nabla^2)$-terms included, with the result
\be \label{a2}
\frac{\partial \epsilon(r,t)}{\partial t} \simeq D_T \nabla^2 \epsilon(r,t)
\ee
with a heat diffusivity
\be \label{a3}
D_T=\frac{1}{2} \lambda_0 a^2 \sum_{m=1}^M m^2
= \frac{1}{12} \lambda_0 a^2 M (M+1) (2M+1)
\ee
It is convenient to eliminate $a$ and $M$ from this expression in
favor of the interaction range $r_c=Ma$ and of the {\it reduced
density} $\rho$. Here $\rho \varpi_d $ ($\varpi_1 = 2$) is
according to \Ref{2.3} the number of particles interacting with a
given particle, which  is here $2 M$. So, $\rho = M$, and the heat
diffusivity becomes,
\be \label{a4}
D_T(\rho) = \frac{1}{6} \rho \lambda_0 r_c^2 \left(1+\frac{1}{\rho}\right)
\left(1+\frac{1}{2\rho}\right).
\ee
In the limit of {\it large density} the heat diffusivity approaches
the value $D_\infty(\rho) = \frac{1}{6} \rho \lambda_0 r_c^2$. In
Fig. 12 the analytic result for $R_1(\rho) = D_T(\rho) /
D_\infty(\rho) = (1+1/\rho)(1+1/2\rho)$ is compared with computer
simulations of the DPD-solid in a one dimensional lattice and shows
excellent agreement. We note that for the case of n.n.
interactions ($\rho=1$) the heat diffusion becomes $D_T(\rho=1) =
\frac{1}{2} \lambda_0 a^2$.

%{\bf Figure 12.}
\begin{figure}[h]
\includegraphics[width=5.5cm,angle=-90]{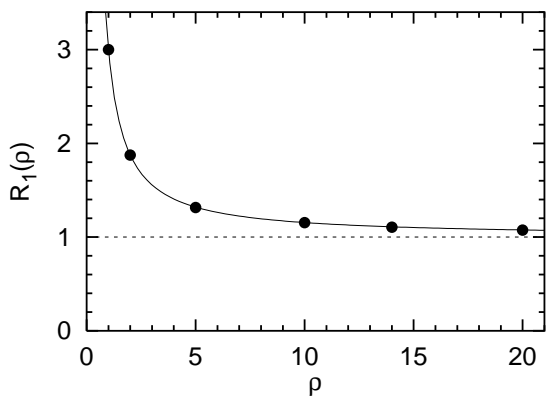}
\caption{A comparison of the simulations
with the theoretical predictions for the heat diffusivity in the
one-dimensional lattice model for
$R_1(\rho)=D_T(\rho)/D_\infty(\rho)=(1+\frac{1}{\rho})
(1+\frac{1}{2\rho})$ with a few simulation data. Note that $\rho$
is an integer. There is excellent agreement.}
\end{figure}

\end{document}